# Real-time quantum feedback prepares and stabilizes photon number states

Clément Sayrin[1], Igor Dotsenko[1], Xingxing Zhou[1], Bruno Peaudecerf[1], Théo Rybarczyk[1], Sébastien Gleyzes[1], Pierre Rouchon[2], Mazyar Mirrahimi[3], Hadis Amini[2], Michel Brune[1], Jean-Michel Raimond[1] & Serge Haroche[1,4]

[1]Laboratoire Kastler Brossel, ENS, UPMC–Paris 6, CNRS, 24 rue Lhomond, 75005 Paris, France.

[2]Centre Automatique et Systèmes, Mathématiques et Systèmes, Mines ParisTech, 60 Boulevard Saint-Michel, 75272 Paris Cedex 6, France.

[3]INRIA Paris-Rocquencourt, Domaine de Voluceau, BP 105, 78153 Le Chesnay Cedex, France.

[4]Collège de France, 11 place Marcelin Berthelot, 75231 Paris Cedex 05, France.

**Feedback loops are at the heart of most classical control procedures. A *controller* compares the *signal* measured by a *sensor* (system output) with the *target* value (setpoint). It adjusts then an *actuator* (system input) in order to stabilize the signal towards its target. Generalizing this scheme to stabilize a micro-system's quantum state relies on *quantum feedback*[1-3], which must overcome a fundamental difficulty: the measurements by the sensor have a random back-action on the system. An optimal compromise employs *weak measurements*[4,5] providing partial information with minimal perturbation. The controller should include the effect of this perturbation in the computation of the actuator's unitary operation bringing the *incrementally perturbed* state closer to the target. While some aspects of this scenario have been experimentally demonstrated for the control of quantum[6-9] or classical[10,11] micro-system variables, continuous feedback loop operations permanently stabilizing quantum systems around a target state have not yet been realized. Following a method inspired by ref. 12 and described in ref. 13, we have implemented such a real-time stabilizing quantum feedback scheme. It prepares on demand photon** **number states (Fock states) of a microwave field in a superconducting cavity C and subsequently reverses the effects of decoherence-induced field quantum jumps[14-16]. The *sensor* is a beam of atoms crossing C which repeatedly performs weak quantum non-demolition measurements of the photon number[14]. The *controller* is implemented in a real-time computer commanding the injection (*actuator*), between measurements, of adjusted small classical fields in C. The microwave field is a quantum oscillator usable as a quantum memory[17] or as a quantum bus swapping information between atoms[18]. By demonstrating that active control can generate non-classical states of this oscillator and combat their decoherence[15,16], this experiment is a significant step towards the implementation of complex quantum information operations.**

A Fock state with $n$ photons is hard to generate and very fragile. Prepared in a cavity of damping time $T_c$, it survives on the average during $T_c/n$ before undergoing a quantum jump towards the $|n-1\rangle$ Fock state. In contrast, classical Glauber states[19], which are coherent superpositions of Fock states with an average photon number $\bar{n}$ and a Poisson photon number probability distribution $P(n) = \exp(-\bar{n})(\bar{n}^n/n!)$, are much easier to prepare and more robust. Glauber states are easily obtained by coupling the initially empty cavity to a classical field source for a fixed amount of time. This operation amounts to the translation of the field in its phase space from the vacuum ($\bar{n} = 0$ coherent state) to a final coherent state having an amplitude $\alpha = \sqrt{n_0}$. After the source is switched off, the field remains a coherent state with an exponentially decaying amplitude, $\bar{n}$ becoming $\bar{n}(t) = \bar{n}_0 \exp(-t/T_c)$.

Experimental methods to prepare Fock states in a cavity C start from a coherent state and exploit the coupling of the field to two-level qubits[14,20,21]. A deterministic procedure feeds quanta one at a time into the field initially in vacuum by swapping its energy with a qubit periodically re-pumped in its excited state[21]. This method, which has been generalized to synthesize arbitrary superpositions of Fock states[22], cannot counteract decoherence because it does not provide real time information on the actual field state in C. Fock states can also be prepared by a quantum non-demolition (QND) measurement performed on an initial coherent state with $\bar{n}_0 \neq 0$ [(14)]. Atomic qubits probe the field one at a time and $n$ is progressively pinned down to an inherently random value, the probability for finding $n$ being the $P(n)$ value of the initial coherent field. This QND method provides real time information about the field state history during the process. This information can be used for a deterministic steering of the field towards a target Fock state $|n_t\rangle$, as well as for detection and subsequent correction of quantum jump events. We have performed a quantum feedback experiment by combining the detection of successive atoms with field phase-space translations of controlled amplitudes. We thus prepare

Fock states $|n_t\rangle$ on demand and, on the average, stabilize them by bringing the field back into them after decoherence-induced quantum jumps.

The experiment is performed in a superconducting cavity C with $T_c = 65$ ms cooled at 0.8 K (see Fig. 1 and Supplementary Methods). It is initially fed by the source S which prepares a coherent state with a real amplitude $\alpha_t = \sqrt{n_t}$. The quantum sensors are circular Rydberg atoms prepared in B at regular $T_a = 82$ µs time intervals[18,23]. The number of Rydberg atoms in each sample obeys a Poisson statistics, with 0.6 atoms per sample on the average. The atomic states $|g\rangle$ and $|e\rangle$ with principal quantum numbers 51 and 50 are the 0 and 1 states of a qubit slightly off-resonant with C (atom-cavity detuning $\delta/2\pi = 245$ kHz). The qubit coherence undergoes in C a light-induced phase-shift linear in the photon number (phase-shift per photon $\phi_0 = 0.256\,\pi$). This phase-shift is measured by a Ramsey interferometer ($R_1$ and $R_2$). Detecting each atomic sample in D provides partial information about the number of photons in C.

Each iteration of the feedback loop[13] consists in a sample detection by the detector D, a cavity field state estimation by the controller K and a field translation performed by the actuator S. In each iteration, K first updates its estimation of the field density operator $\rho$ based on the detection outcome and corrects this estimation by taking into account the effect of cavity relaxation at finite temperature during the iteration time $T_a$. It then computes the amplitude $\alpha$ of the translation described by the operator $D(\alpha) = \exp(\alpha a^\dagger - \alpha^* a))$ ($a$: photon annihilation operator). Since the initial and target density operators are real, we restrict the translations to real $\alpha$'s. The field translation minimizes a proper "distance" $d(\rho_t, D(\alpha)\rho D(-\alpha))$ (defined below) between the displaced state and the target state $\rho_t = |n_t\rangle\langle n_t|$. Finally, at the end of each feedback loop iteration, K calculates the translated field's state which is to be used at the beginning of the next iteration. Note that this quantum state estimation, performed on a single quantum trajectory, cannot be obtained from the measurement data only. It also incorporates all available information on the state preparation, displacements and relaxation.

In an ideal experiment, with exactly one atom prepared and perfectly detected in each sample, a detection in $|e\rangle$ or $|g\rangle$ would actualize the state estimation by the mapping $\rho \rightarrow M_j\rho M_j^\dagger/\text{Tr}(\rho M_j^\dagger M_j)$ ($j=e,g$) with $M_e = \cos[(\phi_r+\phi_0(N+½))/2]$ and $M_g = \sin[(\phi_r+\phi_0(N+½))/2]$ where $\phi_r$ is the tunable phase of the Ramsey interferometer and $N = a^\dagger a$ the photon number operator. This qubit detection is a weak measurement of $N$ associated to the Positive Operator Valued Measure (POVM) $\Pi_j = M_j^\dagger M_j$. In the actual experiment, the measurement-induced state mapping takes into account all known and independently measured imperfections: possibility of 0 and 2 atoms in atomic samples, finite detection efficiency and wrong atomic state assignment (see Supplementary Methods for details). If, for instance, no detection occurs, there is a probability that no atom was present in the sample, in which case the field state does not change. There is another probability that the detector has failed to detect a single qubit, in which case the field should be updated according to the mapping $\rho \rightarrow \Sigma_j M_j \rho M_j^\dagger$. It is also possible that the detector has missed two qubits, in which case the updating would be $\rho \rightarrow \Sigma_{jj'} M_{j'} M_j \rho M_j^\dagger M_{j'}^\dagger$ ($j,j'=e,g$). The probabilities that these situations have occurred, conditioned to the fact that no detection was made, are obtained by a classical Bayesian inference argument. Similar Bayesian reasonings are used to infer the probabilities which affect the mapping when one or two qubits are detected. The state estimation also takes into account the back-action on the field of the yet undetected samples which are on their 344 μs long flight from C to D.

The control law relies on a Lyapunov-based state stabilization[24]. Its efficiency depends upon the definition of the distance $d(\rho_t,\rho)$ (the control Lyapunov function) between the field estimation $\rho$ and the target $\rho_t = |n_t\rangle\langle n_t|$. In the simulations described in ref. 13, the simple definition $d = 1-\langle n_t|\rho|n_t\rangle$ was used. This distance vanishes when the target is reached, but it does not discriminate the $n \neq n_t$ Fock states whose distances to the target are all equal to 1. A better choice defines the distance as $d = 1-\text{Tr}(\Lambda^{(n_t)}\rho)$, where $\Lambda^{(n_t)}$ is a diagonal matrix with $\langle n_t|\Lambda^{(n_t)}|n_t\rangle = 1$ and the other elements $\langle n|\Lambda^{(n_t)}|n\rangle$ ($n \neq n_t$) decreasing monotonically with $|n - n_t|$. In this case, $d$ carries information not only about the probability that the field contains $n_t$ photons, but also about how far from $n_t$ non-negligible $P(n)$ values are found. The $\Lambda^{(n_t)}$ matrix is optimized by performing simulations of feedback trajectories and adjusting the $\Lambda_{nn}^{(n_t)}$ coefficients to obtain the fastest convergence. Based on this value of $\Lambda^{(n_t)}$, K searches, at each iteration step, for the α value which minimizes $d(\rho_t, D(\alpha)\rho D(-\alpha))$. To reduce the computation time, it uses an expansion of $D(\alpha)$ up to second order and determines, under this approximation, an optimal field translation with α in the [−0.1,+0.1] interval (see Supplementary Methods).

Figure 2 shows the experimental records of two 164 ms long feedback sequences aiming at $|n_t = 2\rangle$ (left column) and $|n_t = 3\rangle$ (right column), respectively. The measurement outcomes (Fig. 2a) are fed into K which updates the distance to the target (Fig. 2b) and computes the optimal field translation applied by S (Fig. 2c). This results in the estimated probabilities for finding $n = n_t$, $n < n_t$ and $n > n_t$ number of photons in C (Fig. 2d). After an initial transient period lasting about 20 ms (240 iterations, about 50 detected atoms), the distance to the target drops to a small value and the field reaches $|n_t\rangle$ with a fidelity $\langle n_t|\rho|n_t\rangle \approx 0.8$. The actuator operates during the convergence phase and then quiets down until the field undergoes a quantum jump towards $|n_t-1\rangle$. The distance to the target then features a sudden burst, inducing S to become active again, until the target state is restored, in a time of about 10 to 20 ms (120-240 iterations). Later quantum jumps are corrected in the same way. The rate of quantum jumps increases with $n$, which explains that S is somewhat more active for $n_t = 3$ than for $n_t = 2$, with a slightly reduced average fidelity. Similar recordings obtained for $n_t = 1$ and 4 are shown in Supplementary Methods.

Figure 2e shows snapshots of the density operator $\rho$ as estimated by the feedback controller K. For each sequence, we have represented from left to right the initial coherent state, the states after the convergence has been observed, shortly after a quantum jump has been detected, and finally during the recovery from the jump. Note that the initially large off-diagonal elements $\rho_{nn'}$ ($n \neq n'$) vanish when the field state reaches the target represented by a single peaked diagonal matrix. A quantum jump is detected as a fast increase of the $|n_t-1\rangle$ state probability at the expense of that of the $|n_t\rangle$ state, without build-up of off-diagonal elements. The recovery from the jump is due to small coherent field injections which create transient $\rho_{nn'}$ coherences between Fock states close to $n = n_t$. Supplementary information presents movies featuring the complete evolution of the field density operator during feedback loops.

For each $n_t$ value, we have recorded large sets of feedback trajectories with two different stopping conditions. 4,000 of them are interrupted by the controller at 164 ms as in Fig. 2 (fixed time stop) and about 3,900 when $P(n_t)$ is found by K to be greater than 0.8 in 3 successive iterations (fixed fidelity stop). For each $n_t$ and stopping condition, the final ensemble-averaged photon number distribution $P_{QND}(n)$ is reconstructed independently from the K estimation, using additional probe atoms sent immediately after the interruption of the feedback loop (see Supplementary Methods). The blue and red bars in Fig. 3 give the $P_{QND}(n)$s obtained for the fixed time stop and the fixed fidelity stop, respectively, for $n_t = 1$ to 4. For reference, the green histograms show the measured photon number distribution of the initial coherent state, well described by the Poisson statistics. The high values of the red bars peaking at $n_t$ show the actual fidelity of the state preparation. The blue bar histograms are somewhat broader than the red ones because, on the average, the field resides for fractions of time in states with $n \neq n_t$ due to the finite time it takes to correct a quantum jump. These fixed stop time histograms are however narrower than the initial ones of the coherent field, with $P_{QND}(n_t)$ about 2 times larger than the corresponding value for the coherent state.

We have also analysed the convergence speed towards the target. Figure 4a shows the fraction of trajectories having reached the 0.8 fidelity threshold for $n_t = 3$ as a function of time.

The convergence time (for which 63% of the trajectories have converged) is 50 ms. We compare this result with that of an optimized trial-and-error projection method based on a QND measurement. The photon number of an initial Glauber state with $\sqrt{n_t}$ amplitude is measured by QND probe qubits sent for a fixed time τ. The preparation is declared successful if the inferred probability for $n_t$ is > 0.8. Otherwise, the field is reset to the initial state and the procedure repeated until the threshold is reached. Choosing τ = 14 ms optimizes the convergence rate. The stepped line in Fig. 4a shows that the convergence time is now 250 ms, 5 times longer than that of the quantum feedback method.

We have finally investigated the recovery dynamics from a quantum jump out of $|n_t = 3\rangle$. We prepare the field in the $|n_t -1 = 2\rangle$ Fock state, using a projective QND measurement. We then start a feedback loop with the initial estimated photon number distribution given by the red histogram in Fig. 3c. We thus simulate experimentally the situation in which the field has suddenly jumped in $|n = n_t -1\rangle$ while K still "believes" that $n = n_t$. Figure 4b presents the time evolution of the subsequent $\bar{P}(n,t)$s estimated by K and averaged over 2,561 trajectories. Within about 3 ms (~7 detected atoms), K "realizes" that the jump has occurred (rapid drop of $P(n_t)$ and fast rise of $P(n_t-1)$) and activates the control injection. The field comes back to its steady state (with $\bar{P}(n_t) = 0.43$, this value being limited by subsequent random quantum jumps) within ~15 ms.

We have implemented a real-time quantum feedback procedure generating on demand and stabilizing photon number states by reversing the effects of decoherence-induced quantum jumps. This experiment, which combines quantum measurements and deterministic corrections, presents obvious similarities with quantum error correction codes[25] demonstrated with photons[26], ions[27], spins[28] or superconducting qubits[29]. The long cavity damping time of our cavity QED set-up is an asset since it allows the controller to perform in real time complex estimation and optimization operations. We plan to perform a variant in which the classical actuator source will be replaced by Rydberg atoms delivering single photons in the cavity. The same set-up could also be used to perform adaptive photon number measurements in which the successive qubit settings will be modified in real time, taking into account the results of previous detections[23]. We are also considering applying similar quantum feedback strategies to the stabilization of even more exotic states, such as Schrödinger cat states of radiation[30].

**Acknowledgements** This work was supported by the Agence Nationale de la Recherche (ANR) under the projects QUSCO-INCA, EPOQ2 and CQUID, and by the EU under the IP project AQUTE and ERC project DECLIC.

**Author Contributions** C.S. and I.D. contributed equally to this work. P.R., M.M. and H.A. contributed to the design and optimization of the feedback control.

**Author Information** Correspondence and requests for materials should be addressed to S.H. (haroche@lkb.ens.fr).

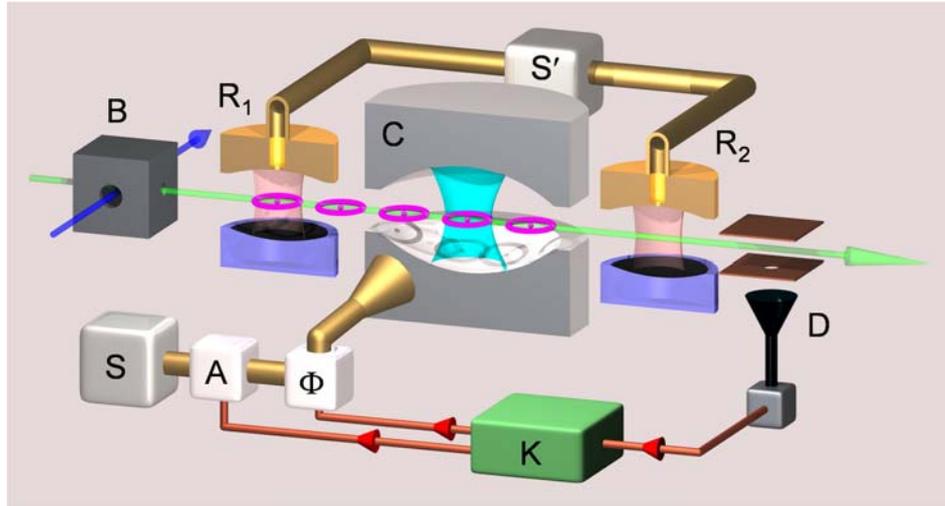

**Figure 1 | Scheme of the quantum feedback set-up.** An atomic Ramsey interferometer (auxiliary cavities $R_1$ and $R_2$) sandwiches the superconducting Fabry-Perot cavity C resonant at 51 GHz and cooled at 0.8 K (mean number of blackbody photons: 0.05). The pulsed classical source S' induces $\pi/2$ pulses resonant with the $|g\rangle \rightarrow |e\rangle$ transition in $R_1$ and $R_2$ (with relative phase $\phi_r$) on the velocity selected (v = 250 m/s) Rydberg atom qubits prepared by laser excitation from a Rubidium atomic beam in B. The field-ionization detector D measures the qubits in the *e*/*g* basis with a 35% detection efficiency and a few percent error rate (see Supplementary Methods). The actuator S feeds C by diffraction on the mirror edges. The controller K (CPU-based ADwin Pro-II system) collects information from D to determine the real translation amplitude $\alpha$ applied by S. It sets the S-pulse duration through a pin-diode switch A (63 μs pulse for $|\alpha| = 0.1$) as well as a 180° phase-shifter $\Phi$ controlling the sign of $\alpha$.

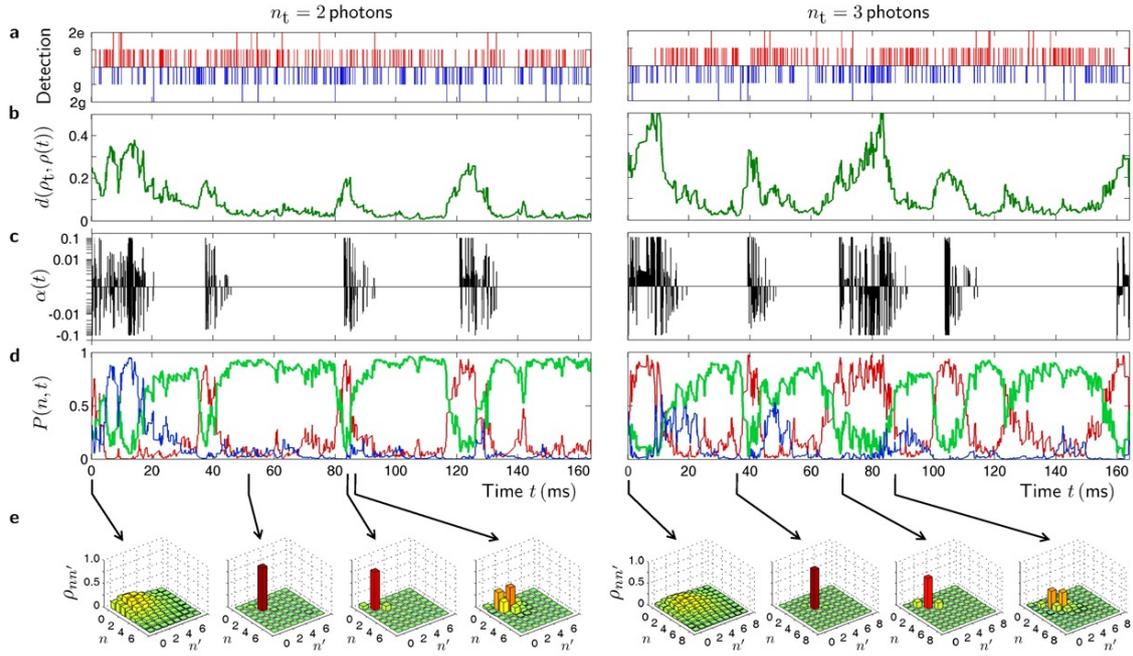

**Figure 2 | Individual quantum feedback trajectories.** Two feedback runs lasting 164 ms (2,000 loop iterations) stabilizing $|n_t = 2\rangle$ (left column) and $|n_t = 3\rangle$ (right column). The phase-shift per photon $\phi_0 = 0.256\,\pi$ allows K to discriminate $n$ values between 0 and 7. For $n_t = 2$, the Ramsey phase is $\phi_r = -0.44$ rad, corresponding to nearly equal $e$ and $g$ detection probabilities when $n = 2$. For $n_t = 3$, two Ramsey phases $\phi_{r,1} = -0.44$ rad and $\phi_{r,2} = -1.24$ rad are alternatively used, corresponding to equal $e$ and $g$ probabilities when $n = 2$ and $n = 3$, respectively.
**a**, Sequences of qubit detection outcomes. The detection results are shown as blue downwards bars for $g$ and red upwards bars for $e$. Two-atom detections appear as double length bars.
**b**, Estimated distance between the target and the actual state. **c**, Applied $\alpha$-corrections (shown in log-scale as $\text{sgn}(\alpha)\log|\alpha|$). **d**, Photon number probabilities estimated by K: $P(n = n_t)$ is in green, $P(n < n_t)$ in red, $P(n > n_t)$ in blue. **e**, Field density operators $\rho$ in a Fock-state basis estimated by K at four different times marked by arrows.

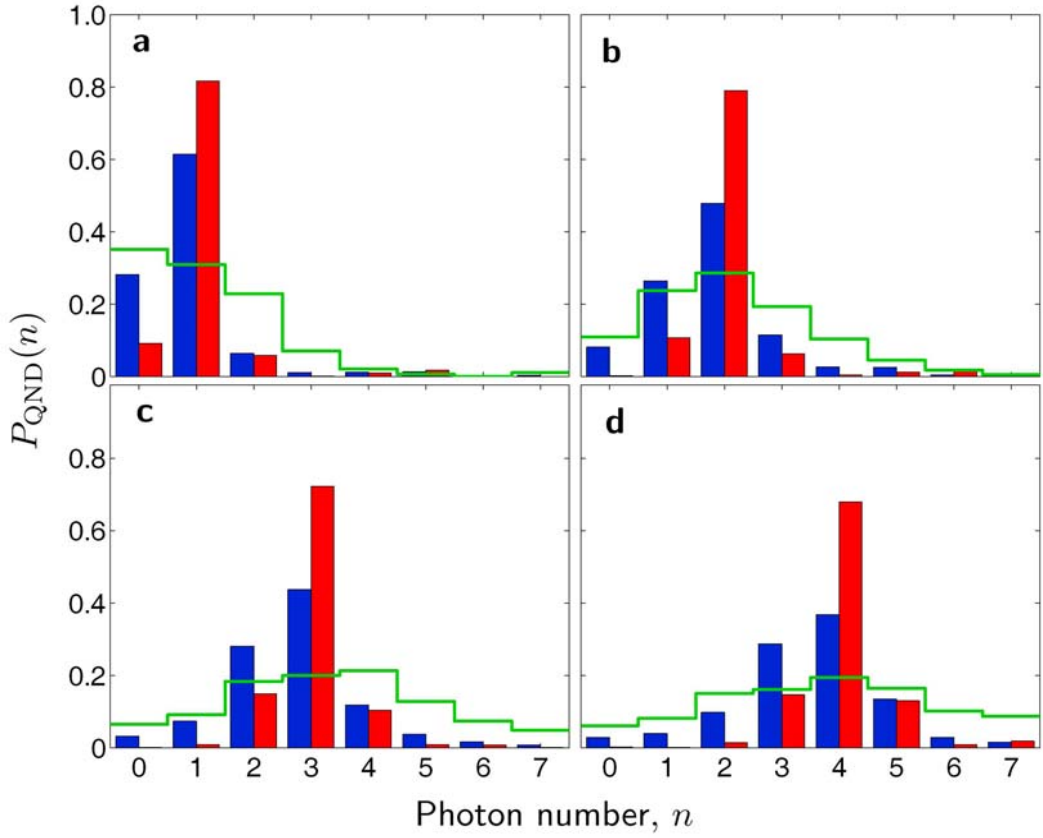

**Figure 3 | Photon number histograms following quantum feedback iterations.**
Plots **a**, **b**, **c**, and **d** correspond to the target photon number states $n_t$ = 1, 2, 3 and 4, respectively. The red histograms correspond to about 3,900 trajectories stopped when $P(n_t)$ has reached for three successive iterations the threshold value 0.8. These histograms describe the field at the time when the controller K has certified the "success" of the quantum feedback procedure. The blue histograms correspond to 4,000 trajectories stopped at a fixed 164 ms time and describe the feedback procedure steady-state. These histograms are reconstructed by a method independent from the feedback estimator. After interrupting the feedback, we record ten additional QND qubit samples (~2 detected atoms) with a Ramsey interferometer phase $\phi_r$ chosen in sequence among 4 values ($\phi_r$ = 1.17, 0.36, −0.44 and −1.24 rad). From these additional qubit detections, we reconstruct the final $P_{QND}(n)$ distribution for each ensemble of trajectories by a maximum likelihood algorithm. Statistical error of the reconstructed $P_{QND}(n)$ for different target states is about 0.01-0.02 for $n = n_t$ and $n_t \pm 1$, and it is significantly smaller than 0.01 for other photon numbers (see Supplementary Methods). The green histograms give the initial coherent state photon number distributions (similar reconstruction performed with a fixed time stop immediately after initial field injection).

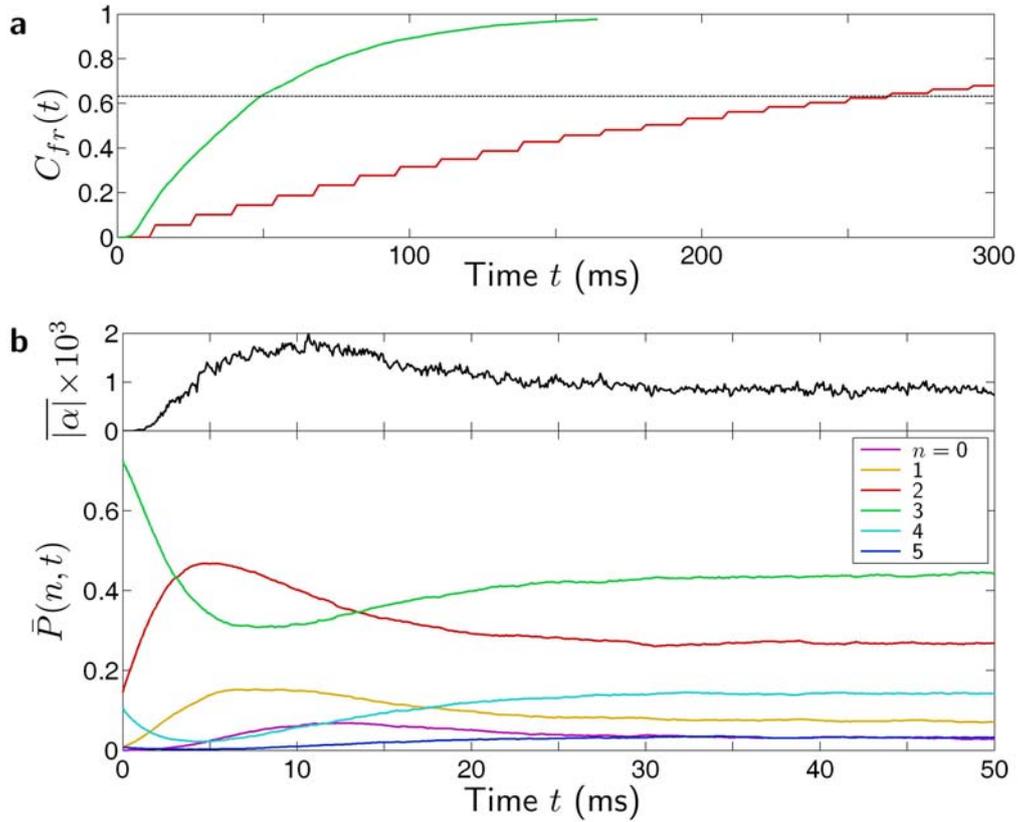

**Figure 4 | Performance of the quantum feedback procedure. a**, Time evolution of the fraction of individual field trajectories $C_{fr}(t)$ having converged towards $|n_t = 3\rangle$ in quantum feedback sequences (smooth line) and in passive QND "trials" (stepped line). Statistics performed over 4,000 and 2,131 trajectories, respectively. The same Ramsey phase settings as in Fig. 2 have been used for both feedback and QND sequences. **b**, Recovery from a quantum jump: the lower plot shows probabilities $\bar{P}(n,t)$ estimated by K and averaged over 2,561 trajectories, following the preparation at $t = 0$ of the Fock state $|n = 2\rangle$ by a QND measurement of an initial coherent state (colour code for the different $\bar{P}(n,t)$s in inset). The Ramsey phase settings are the same as in Fig. 2 for $n_t = 3$. The initial field density matrix of the field estimation algorithm is diagonal and corresponds to the red histogram in Fig. 3c. The experiment thus simulates the reaction of the quantum feedback procedure to a $|3\rangle \rightarrow |2\rangle$ quantum jump occurring at $t = 0$, after the field has converged to the target. The upper plot in **b** shows the variation of the average modulus of the injection amplitude $\overline{|\alpha(t)|}$. Initially zero, $\overline{|\alpha|}$ grows rapidly to a maximum while the quantum jump is reversed. The controller finally quiets and $\overline{|\alpha|}$ returns to its average steady-state value.